\documentclass[12pt]{article}
\usepackage{amsmath,amssymb}
\usepackage{bm}
\usepackage{mathrsfs}
\usepackage{fourier}
\usepackage{multirow}
\allowdisplaybreaks[4]
\newcommand{{\Slashp}}{p\!\!\!\!\!\big/}
\newcommand{{\Slashq}}{q\!\!\!\!\!\big/}
\usepackage[usenames]{color}

\begin{document}

\title{Naturalness, Conformal Symmetry and Duality}

\author{
Yoshiharu \textsc{Kawamura}\footnote{E-mail: haru@azusa.shinshu-u.ac.jp}\\
{\it Department of Physics, Shinshu University, }\\
{\it Matsumoto 390-8621, Japan}\\
}

\date{
August 23, 2013}

\maketitle
\begin{abstract}
We reconsider the naturalness from the viewpoint of effective field theories,
motivated by the alternative scenario that the standard model holds 
until a high-energy scale such as the Planck scale.
We propose a calculation scheme of radiative corrections 
utilizing a hidden duality,
in the expectation that the unnaturalness for scalar masses
might be an artifact in the effective theory
and it could be improved if features of an ultimate theory are taken in.
\end{abstract}


\section{Introduction}
\label{Introduction}

It might be a good time to reconsider various concepts 
concerning the Higgs boson mass $m_h$ 
and the physics beyond the standard model (SM),
on the basis of recent experimental results at LHC.

The discovery of the Higgs boson at LHC 
with the observed value $m_h \doteqdot 126$GeV~\cite{LHC1,LHC2} is quite suggestive.
It rekindles the question whether $m_h$ is a natural parameter 
or not~\cite{Susskind,tHooft,Veltman}. 
Furthermore, evidences of new physics 
such as supersymmtry (SUSY), compositeness and extra dimensions 
have not yet been discovered,
and this fact can be the turning point of the particle physics,
because the gauge hierarchy problem~\cite{GH1,GH2} would be revisited.

Therefore, it is interesting to reexamine the validity of concepts
relating $m_h$ and the physics beyond the SM, from various aspects.
In this paper, we reconsider the naturalness of $m_h$
from the viewpoint of effective field theories including the SM.
Our study is motivated by the alternative scenario 
that the SM (modified with massive neutrinos) holds 
until a high-energy scale such as the Planck scale $M_{\rm Pl}$~\cite{Shaposhnikov,Nielsen}.
We expect that the unnaturalness for scalar masses
might be an artifact in the effective theory,
and it could be improved if features of an ultimate theory are taken in
and ingredients of the effective theory are enriched.
We reanalyze radiative corrections on scalar masses using the $\phi^4$ theory,
and propose a calculation scheme utilizing a hidden duality.

The outline of this paper is as follows.
In the next section, we review the naturalness and its relevant symmetries.
We give a suggestion for the subtraction of quadratic divergences
by presenting a calculation scheme, 
in consideration of a duality relating integration variables, in Sect. 3.
In the last section, we give conclusions and discussions.

\section{Naturalness and conformal symmetry}
\label{Naturalness and conformal symmetry}

Let us first recall the concept of naturalness.
According to 't Hooft~\cite{tHooft}, the naturalness is based on the dogma that
{\it $\lq\lq$at any energy scale $\mu$, a physical parameter 
or set of physical parameters $a_i(\mu)$ is allowed to be very small 
only if the replacement $a_i(\mu) = 0$ would increase
the symmetry of the system.}''
We refer to this type of parameter as a natural parameter.

We discuss the naturalness of fermion masses and scalar masses,
from the viewpoint of low-energy effective theories 
such as the quantum electrodynamics (QED), the $\phi^4$ theory and the SM.

\subsection{Naturalness of fermion masses}
\label{Naturalness and fermion masses}

The electron mass $m_{\rm e}$ is listed as a natural parameter
in the QED.
When we set $m_{\rm e} =0$, the classical global chiral symmetry appears.
Here, the chiral symmetry is the invariance of action integral
under the different phase changes for Weyl fermions $\psi_L$ and $\psi_R$ 
with the different chiralities, i.e.,
$\psi_L \to e^{i\theta_L} \psi_L$ and $\psi_R \to e^{i\theta_R} \psi_R$, $(\theta_L \ne \theta_R)$.
Note that the chiral symmetry is broken down, even in the massless case $m_{\rm e} =0$, such that
\begin{eqnarray}
\langle \partial_{\mu} (\psi^{\dagger}_L \overline{\sigma}^{\mu} \psi_L) \rangle 
= \frac{e^2}{32 \pi^2} \varepsilon_{\mu\nu\alpha\beta} F^{\mu\nu} F^{\alpha\beta}~,~~
\langle \partial_{\mu} (\psi^{\dagger}_R {\sigma}^{\mu} \psi_R) \rangle 
= - \frac{e^2}{32 \pi^2} \varepsilon_{\mu\nu\alpha\beta} F^{\mu\nu} F^{\alpha\beta}~,
\label{jLR}
\end{eqnarray}
in the presence of the axial $U(1)$ anomaly.
In the massive case, the $U(1)$ vector current defined as
$j_V^{\mu} = \overline{\psi} \gamma^{\mu} \psi
= \psi^{\dagger}_L \overline{\sigma}^{\mu} \psi_L+\psi^{\dagger}_R {\sigma}^{\mu} \psi_R$
is conserved,
and the $U(1)$ axial vector current defined as
$j_A^{\mu} = \overline{\psi} \gamma_5 \gamma^{\mu} \psi
= \psi^{\dagger}_L \overline{\sigma}^{\mu} \psi_L-\psi^{\dagger}_R {\sigma}^{\mu} \psi_R$
is anomalous such that
\begin{eqnarray}
\langle \partial_{\mu} j_A^{\mu} \rangle 
= 2i(m_{\rm e} + \delta m_{\rm e}) (\psi^{\dagger}_L \psi_R - \psi^{\dagger}_R \psi_L)
+ \frac{e^2}{16 \pi^2} \varepsilon_{\mu\nu\alpha\beta} F^{\mu\nu} F^{\alpha\beta}~,
\label{jA}
\end{eqnarray}
where $\delta m_{\rm e}$ represents the radiative corrections on the tree level mass $m_{\rm e}$.

The $\delta m_{\rm e}$ at the one-loop level is given by
\begin{eqnarray}
\delta m_{\rm e} = \frac{3 \alpha}{4\pi} m_{\rm e}
\left(\ln \frac{\Lambda^2}{m_{\rm e}^2} + \frac{1}{2}\right)~,
\label{deltame}
\end{eqnarray} 
where $\alpha \equiv e^2/(4\pi)$ and
$\Lambda$ is a cutoff scale.
In the limit of $m_{\rm e} \to 0$, $\delta m_{\rm e}$ also vanishes.
This feature holds for a higher order corrections,
and the chiral symmetry is not broken down perturbatively
(although it is broken down anomalously without threatening the consistency of theory).
Hence, the chiral symmetry is regarded as a powerful concept to control quantum corrections.

A classical conformal symmetry also appears in the limit of $m_{\rm e} \to 0$,
and is broken down in the presence of anomaly.
For instance, the scale invariance is broken down as
\begin{eqnarray}
\langle {T^{\mu}}_{\mu} \rangle 
= (m_{\rm e} + \delta m_{\rm e}) (\psi^{\dagger}_L \psi_R + \psi^{\dagger}_R \psi_L)
+ \frac{\beta_{\alpha}}{\alpha} F^{\mu\nu} F_{\mu\nu}~,
\label{T-QED}
\end{eqnarray} 
where $\beta_{\alpha}$ is the $\beta$ function for $\alpha$.
In the QED, the conformal symmetry seems to play a same role as the chiral symmetry does.

In the SM, the chiral symmetry has a superior quality to the conformal symmetry,
because the chiral symmetry such as $SU(2)_L \times U(1)_Y$ 
becomes a local one and is not broken down
either perturbatively or anomalously, and
the conformal symmetry is broken down 
not only with the negative mass squared of Higgs doublet
explicitly, but also in the presence of anomalous terms.
The chiral gauge symmetry is broken down spontaneously with the vacuum expectation value of
the Higgs boson $v=246$GeV,
and fermions $\psi_f$ acquire masses $m_f = y_f v/\sqrt{2}$ via the Higgs mechanism,
where $y_f$ are Yukawa coupling constants.
On the other hand, the global chiral symmetry enhances in the limit of $y_f \to 0$.
In this way, the smallness of fermion masses, 
comparing with a high-energy scale $M_{\rm U}$ 
such as the gravitational scale $M \equiv M_{\rm Pl}/\sqrt{8\pi} = 2.4 \times 10^{18}$GeV,
stems from the smallness of $v$, comparing with $M_{\rm U}$.
Furthermore, the smallness of fermion masses except for the top quark mass, 
comparing with the weak gauge boson mass $M_W = g v/2$,
originates from the smallness of $y_f$, comparing with the $SU(2)_L$ gauge coupling constant $g$.
Hence, it is considered that the chiral symmetry is responsible for 
the smallness of SM fermion masses.

\subsection{Naturalness of scalar masses}
\label{Naturalness and scalar masses}

Next, we study the relation between the relevant symmetry of scalar mass $m_{\phi}$
and radiative corrections on $m_{\phi}$, 
in order to take a hint for a naturalness of $m_h$.
Unless a theory has dimensional parameters except for $m_{\phi}$,
the classical conformal symmetry appears in the limit of $m_{\phi} \to 0$.
In the same way as the QED, the scale invariance is broken down as
\begin{eqnarray}
\langle {T^{\mu}}_{\mu} \rangle 
= (m_{\phi}^2 + \delta m_{\phi}^2) \phi^2 + \sum_{k} \beta_k \mathcal{O}_k~,
\label{T}
\end{eqnarray} 
where $\delta m_{\phi}^2$ represents the radiative corrections on $m_{\phi}^2$,
$\beta_k$ are $\beta$ functions for coupling constants $a_k$,
and $\mathcal{O}_k$ are operators with the mass dimension 4.

In the $\phi^4$ theory,
$\delta m_{\phi}^2$ at the one-loop level is most commonly written by
\begin{eqnarray}
\delta m_{\phi}^2 = \frac{\lambda_{\phi}}{32\pi^2}
\left(\Lambda^2 - m_{\phi}^2 \ln \frac{\Lambda^2}{m_{\phi}^2}\right) + \cdots~,
\label{deltamphi}
\end{eqnarray} 
where $\lambda_{\phi}$ is the quartic self-coupling constant of $\phi$,
and the ellipsis stands for $\Lambda$ independent terms.

For example, the unregularized one is given by
\begin{eqnarray}
\delta m_{\phi}^2 = \frac{\lambda_{\phi}}{2} \int_{-\infty}^{\infty} \frac{d^4p}{(2\pi)^4}
\frac{1}{p^2 + m_{\phi}^2}
= \frac{\lambda_{\phi}}{32\pi^2} \left(\int_{0}^{\infty} dp^2 
+ \int_{0}^{\infty} \frac{-m_{\phi}^2}{p^2 + m_{\phi}^2} dp^2\right)~,
\label{deltamphi-un}
\end{eqnarray}
where we rotate to the Euclidean space and carry out the integration for the angles of momentum space.
The $\delta m_{\phi}^2$ of (\ref{deltamphi}) is obtained by replacing $\infty$ by $\Lambda^2 - m_{\phi}^2$
on the final expression in (\ref{deltamphi-un}).

As seen from (\ref{deltamphi}), it is widely thought that 
$m_{\phi}$ is not a natural parameter,
because $\delta m_{\phi}^2$ does not vanish in the limit of $m_{\phi}^2 \to 0$,
in the appearance of the quadratic term of $\Lambda$.

However, if the quadratic term is subtracted or absent from some reason,
$m_{\phi}$ can be a natural parameter.
Bardeen reexamined the naturalness in the SM
and pointed out that the classical scale invariance implies 
the naturalness of the Higgs boson mass $m_h$~\cite{Bardeen}.
The reasoning is illustrated as follows.
In the SM, the scale invariance is broken as
\begin{eqnarray}
\langle {T^{\mu}}_{\mu} \rangle 
= (m_h^2 + \delta m_h^2) |H|^2 + \sum_{k} \beta_k \mathcal{O}_k~,
\label{T-H}
\end{eqnarray} 
where $\delta m_h^2$ represents the radiative corrections on $m_h^2$.
The anomalous terms are quantum corrections induced from loop contributions 
due to particles with masses smaller than the reference energy scale.
It is quite unlikely that the radiative corrections on masses affect them.
Hence, the anomalous divergence of the scale current should remain in the limit of $m_h \to 0$,
and $\delta m_h^2$ should be proportional to not $\Lambda^2$ but $m_h^2$.
In other words, the classical symmetries should be restored in the limits of $m_h \to 0$
and $\beta_{k} \to 0$.

In effective field theories, ambiguities can exist in the regularization procedure,
and such ambiguities, in most cases, are resolved by considering
symmetries realized manifestly at the low-energy scale~\cite{Jackiw}.
If we had a theory with a high calculability and predictability,
regularization dependent quantities would be absent.
In this regard, quantities depending on the regularization method
should be subtracted or eliminated,
unless the subtraction induces any physical effects.

The dimensional regularization is known as a regularization procedure, 
that does not induce quadratic divergences for scalar masses.
Using it, $\delta m_{\phi}^2$ at the one-loop level is given by
\begin{eqnarray}
\delta m_{\phi}^2 = \frac{\lambda_{\phi}}{32\pi^2} 
m_{\phi}^2\left(-\frac{2}{\epsilon} + \gamma - 1\right) + \cdots~,
\label{deltamphi-dim}
\end{eqnarray}
where $\epsilon = 4-D$ ($D$ is the dimension of space-time)
and $\gamma = 0.577\cdots$ is the Euler constant.
The $\delta m_{\phi}^2$ becomes infinite in the limit of $\epsilon \to 0$, i.e., $D \to 4$.
The $2/\epsilon$ corresponds to $\ln ({\Lambda^2}/{m_{\phi}^2})$,
and then the quadratic divergence is absent.

Fujikawa gave a scheme on the subtractive renormalization of 
the quadratic divergences of scalar mass~\cite{Fujikawa}.
In case that the subtraction of quadratic divergences 
induces no physical effects on the low-energy theory,
such a scheme is useful to treat physical quantities including scalar masses.

Aoki and Iso studied the quadratic divergences of scalar mass 
from the viewpoint of the Wilsonian renormalization group, and
found that they can be absorbed into a position of the critical surface,
which means the subtraction of them~\cite{A&I}.

Extensions of the SM have been proposed
by adopting the classical conformal invariance as a guiding principle~\cite{M&N,FKM&V,H&K,IO&O,I&O}.\footnote{
A model that both the Planck scale and the weak scale emerge as quantum effects has been proposed~\cite{FKM&V}.
As an extension including dark matter candidates,
a model with a strongly interacting hidden sector, to trigger the breakdown of electroweak symmetry, 
has been constructed~\cite{H&K}.
Recently, various models, to generate the weak scale
and provide dark matter candidates, have been proposed~\cite{HRRS&T,H&S,C&R,FH&R}.
}

\subsection{Naturalness of Higgs boson mass}
\label{Naturalness and Higgs boson mass}

Before we study the subtraction of quadratic divergences
from the viewpoint of hidden symmetry,
we discuss the naturalness of Higgs boson mass.
In the SM,
the radiative corrections on the Higgs mass squared $m_h^2$ at the one-loop level are given by
\begin{eqnarray}
\delta m_h^2 = c_h \Lambda^2 + c'_h m_h^2 \ln \frac{\Lambda^2}{m_h^2} + \cdots~,
\label{deltamh}
\end{eqnarray} 
where $c_h$ and $c'_h$ are functions of the SM parameters such that
\begin{eqnarray}
c_h = \frac{1}{16 \pi^2} \left(6\lambda + \frac{9}{4} g^2 + \frac{3}{4} g'^2 - 6 y_t^2\right)~,~~
c'_h = \frac{1}{16 \pi^2} \left(6\lambda - \frac{9}{4} g^2 - \frac{3}{4} g'^2 + 3 y_t^2\right)~.
\label{c'h}
\end{eqnarray}
Here $\lambda$ is the quartic self-coupling constant of Higgs boson,
$g'$ is the $U(1)_Y$ gauge coupling constant, $y_t$ is the top Yukawa coupling constant,
and contributions from other fermions are omitted.

If we face the quadratic divergences squarely, 
the fine tuning among parameters is necessary to explain 
the observed value $m_h \doteqdot 126$GeV, unless $\Lambda \le O(1)$TeV or $c_h = 0$ is realized.
Here, the condition $\Lambda \le O(1)$TeV means 
that a new physics beyond the SM must exist around the terascale,
unless nature requires the fine tuning.
The condition $c_h = 0$ is equivalent to the Veltman condition 
$m_h^2 = 4 m_t^2 - 2 M_W^2 - M_Z^2$~\cite{Veltman},\footnote{
The same type of condition was derived through a tadpole diagram concerning the Higgs boson~\cite{D&P}.
}
which leads to a value $m_h \doteqdot 320$GeV at the weak scale.\footnote{
Recently, it is pointed out that $c_h = 0$ holds around $M_{\rm Pl}$
and there is a possibility that the bare Higgs mass vanishes there~\cite{HK&O}.
}

If all quadratic divergences are subtracted, $\delta m_h^2$ also vanishes in the limit of $m_h \to 0$.
Then, the classical conformal symmetry seems to control quantum corrections
as the chiral symmetry does.
If this feature holds for a higher order corrections,
the classical conformal symmetry is not broken down perturbatively
(although it is broken down anomalously without threatening the consistency of theory).
In this way, the conformal symmetry might be responsible to
the smallness of Higgs boson mass, comparing with a high-energy scale $M_{\rm U}$.
Or it might be said that the regularization ambiguities can be resolved by the conformal symmetry.

Hence, the problem {\it whether the weak scale relating the Higgs boson mass 
is stabilized against radiative corrections
in the framework of SM} (a narrow definition of the naturalness problem)\footnote{
We use the terminology $\lq\lq$the naturalness problem" in a wider sense,
that is, it should be regarded as a collective term for a fine tuning problem 
concerning mass parameters of scalar fields such as the Higgs mass, 
which contains the gauge hierarchy problem.
}
can be solved by the subtraction of quadratic divergences.
Then, the naturalness can become a powerful guiding principle 
to construct an effective theory.\footnote{
Wells presented an interesting observation toward the SM from the QED
using the naturalness as the guiding principle~\cite{Wells}.
}
In other words, symmetries such as the chiral symmetry, the gauge symmetry and the conformal symmetry
become powerful tools for a realistic model-building, from the viewpoint of the effective field theory.
There is a possibility that all fields, in our low-energy world, are massless at $M_{\rm U}$.

At this stage, the following questions (other parts of the naturalness problem) arise.

One is 
{\it what induces the negative mass squared of the Higgs boson around the weak scale,
starting the massless state at $M_{\rm U}$, this is, what is the origin of the weak scale.}
A possible solution has been proposed based on the extension of the SM
with the $U(1)_{B-L}$ gauge symmetry and new particles around the terascale~\cite{M&N,IO&O}.
In particular, the TeV scale $B-L$ model proposed in \cite{I&O}
has several excellent features such as the classical conformality,
the flatness of Higgs potential at $M_{\rm U}$, 
and the predictability relating $m_h \doteqdot 126$GeV.

The other is the problem {\it whether the weak scale is stabilized against 
large radiative corrections due to heavy particles
in the framework of field theory including a high-energy physics, 
e.g., a grand unified theory.} 
This is (the technical side of) the gauge hierarchy problem~\cite{GH1,GH2}.
For instance, in the presence of heavy particles with masses $M_I$
and some SM gauge quantum numbers,
$m_h^2$ generally receives large radiative corrections of $O(M_I^2)$
in addition to the quadratic term of $\Lambda$ such that
\begin{eqnarray}
\delta m_{h}^2 = \tilde{c}_h \Lambda^2 + c'_{h} m_{h}^2 \ln \frac{\Lambda^2}{m_{h}^2}
+ \sum_{I} c''_{hI} M_I^2 \ln \frac{\Lambda^2}{M_I^2} + \cdots~,
\label{deltamhMI}
\end{eqnarray}
and the stability of the weak scale is threatened.
Here, $\tilde{c}_h$ and $c''_{hI}$ are also functions of parameters.
Then, the fine tuning is indispensable for $M_I^2 \gg m_h^2$
in the appearance of the quadratic term of $M_I$ (a part of logarithmic divergences),
even if the quadratic divergences of $O(\Lambda^2)$ are removed and
unless some miraculous cancellation mechanism works among corrections of heavy particles.
We will come back to this problem in subsection \ref{Different choice}.


\section{Naturalness and duality}
\label{Naturalness and duality}

Let us explore the possibility that the quadratic divergences are removed,
in the expectation that 
{\it the quadratic divergences might be artifacts of regularization procedure
and the calculation scheme can be selected by the physics}.

Concretely, we pursue other reasoning to suggest the subtraction of quadratic divergences,
based on the conjecture that {\it an ultimate theory does not induce 
any large radiative corrections for low-energy fields
owing to a symmetry, and such a symmetry is hidden in the SM.}\footnote{
This conjecture corresponds to one of the guiding principles
to solve the gauge hierarchy problem and the cosmological constant problem,
without SUSY and extra dimensions, proposed by Dienes~\cite{Dienes}.
}
We present a (tricky) calculation scheme, that rules out quadratic divergences
thanks to a hidden duality.

In the following, it is shown that the logarithmic corrections on a scalar mass 
can be picked out by specifying the duality in the effective field theory.

\subsection{Basic idea}
\label{Basic idea}

Our method is based on the following assumptions relating features of an underlying theory.\footnote{
Our idea is inspired by the world-sheet modular invariance in string theory.
We will comment on the world-sheet modular invariance
in subsection \ref{Different choice}.
}\\
(a) There is an ultimate theory, which has a fundamental energy scale.
We denote the scale as $\Lambda$, for simplicity.\\
(b) The ultimate theory has a duality 
between the physics at a higher-energy scale ($E \gtrsim \Lambda$) 
and that at a lower-energy scale ($E \lesssim \Lambda$).
It consists of the following two features.\\
(b1) The physics is invariant under a duality transformation, e.g., $E \to E' = \Lambda^2/E$.\\
(b2) The physics is only described by one of the two energy regions,
relating with each other by the transformation.\\
(c) A remnant of the duality is hidden in quantities of
the low-energy physics involved with $\Lambda$, e.g., radiative corrections on parameters.

To illustrate our idea, let us consider quantum corrections 
on a parameter $a$ at the one-loop level given by,
\begin{eqnarray}
\delta a = \int_{0}^{\infty} f(p^2) dp^2~,
\label{deltaa}
\end{eqnarray}
where $p^2$ is an Euclidean momentum squared for a massless virtual particle
running in the loop, and $f(p^2)$ is a function of $p^2$.

In case that $\delta a$ diverges, the infinities come from $p^2 = \infty$ and/or $p^2 = 0$,
and hence it is ordinarily regularized as
\begin{eqnarray}
\delta a = \int_{\mu_0^2}^{\Lambda^2} f(p^2) dp^2~,
\label{deltaa-reg}
\end{eqnarray}
where $\mu_0$ is a fictitious mass parameter.

Here, let us show that the expression (\ref{deltaa-reg}) is necessarily obtained
and the form of $\delta a$ is restricted, based on the above assumptions,
by specifying the duality transformation.

First, we rewrite (\ref{deltaa}) as
\begin{eqnarray}
\delta a = \int_{\mu_0^2}^{\Lambda^2} f(p^2) dp^2 + \int_{\Lambda^2}^{\Lambda^4/\mu_0^2} f(p^2) dp^2~.
\label{deltaa-2}
\end{eqnarray} 
Note that (\ref{deltaa-2}) is reduced to (\ref{deltaa}) in the limit of $\mu_0^2 \to 0$.
Using the assumption (b), (\ref{deltaa-reg}) is obtained,
if the domain of integration $[\mu_0^2, \Lambda^2]$ is transformed into $[\Lambda^2, \Lambda^4/\mu_0^2]$
under a remnant of duality and the following relation holds,
\begin{eqnarray}
\int_{\mu_0^2}^{\Lambda^2} f(p^2) dp^2 = \int_{\Lambda^2}^{\Lambda^4/\mu_0^2} f(p^2) dp^2~.
\label{r1}
\end{eqnarray}
 
Next, we take $p^2 \to p'^2 = \Lambda^4/p^2$ as the remnant of duality transformation.
Hereafter, we refer to the remnant of duality transformation as the duality transformation
or the duality, in most cases.
Then, using assumptions (b1) and (c), the the following relation is derived, 
\begin{eqnarray}
\int_{\mu_0^2}^{\Lambda^2} f(p^2) dp^2 = \int_{\Lambda^4/\mu_0^2}^{\Lambda^2} f(p'^2) dp'^2
= \int_{\Lambda^2}^{\Lambda^4/\mu_0^2} f(\Lambda^4/p^2) \frac{\Lambda^4}{p^4} dp^2~.
\label{duality}
\end{eqnarray}
From (\ref{r1}) and (\ref{duality}), the form of $f(p^2)$ is restricted as $f(p^2)=c(p^2)/p^2$
where $c(p^2)$ is a function invariant under the change $p^2 \to \Lambda^4/p^2$, e.g.,
$c(p^2) = p^2 + \Lambda^4/p^2$.
Unless we consider effects of heavy particles with masses of $O(\Lambda)$ such as threshold corrections,
$f(p^2)$ does not contain $\Lambda$ and then $\delta a$ is determined as
\begin{eqnarray}
\delta a = c_{-1} \ln \frac{\Lambda^2}{\mu_0^2}~,
\label{deltaa-3}
\end{eqnarray}
where $c_{-1}$ is a $p^2$-independent quantity.

Our procedure can be regarded as not a mere regularization 
but a recipe to obtain finite physical values, because $\Lambda$ is (big but) finite and
infinities are taken away by the symmetry relating integration variables, like string theory.
It is also regarded as the operation to pick out parts that satisfy assumptions.
In case that $f(p^2)$ does not contain $\Lambda$,
it is simply denoted by
\begin{eqnarray}
\delta a = {\rm Du}\left[\int_{0}^{\infty} f(p^2) dp^2\right] 
= {\rm Du}\left[\int_{0}^{\infty} \sum_n c_n \left(p^2\right)^n dp^2\right]
= c_{-1} \ln \frac{\Lambda^2}{\mu_0^2}~,
\label{deltaa-Du}
\end{eqnarray}
where ${\rm Du}[*]$ represents the operation, and $f(p^2)$ is expanded in a series.

\subsection{Radiative corrections on scalar mass}
\label{Radiative corrections on scalar mass}

We apply our method to radiative corrections on $m_{\phi}^2$.

In case that the bare mass is zero,
the unregularized one is given by
\begin{eqnarray}
\delta m_{\phi}^2 = \frac{\lambda_{\phi}}{2} \int_{-\infty}^{\infty} \frac{d^4p}{(2\pi)^4}
\frac{1}{p^2}
= \frac{\lambda_{\phi}}{32\pi^2} \int_{0}^{\infty} dp^2~.
\label{deltamphi-0}
\end{eqnarray}
If we demand that the duality $p^2 \to \Lambda^4/p^2$ is hidden in $\delta m_{\phi}^2$
and the physics can be described by the region below $\Lambda$,
$\delta m_{\phi}^2$ turns out to be zero such that
\begin{eqnarray}
\delta m_{\phi}^2 
= {\rm Du}\left[\frac{\lambda_{\phi}}{32\pi^2} \int_{0}^{\infty} dp^2\right] = 0~.
\label{deltamphi-0-Du}
\end{eqnarray}

Next, we study the case with a non-zero bare mass,
based on the momentum cutoff method and the proper time method.

~~\\
(i) The momentum cutoff method

First, we separate the original one 
into the quadratic and logarithmic divergent parts such that
\begin{eqnarray}
\delta m_{\phi}^2 = \frac{\lambda_{\phi}}{32\pi^2} \int_{0}^{{\Lambda}_{\phi}^2} dp^2 
- \frac{\lambda_{\phi}m_{\phi}^2}{32\pi^2} \int_{0}^{{\Lambda}_{\phi}^2} \frac{dp^2}{p^2 + m_{\phi}^2}~,
\label{deltamphi-sep}
\end{eqnarray}
where ${\Lambda}_{\phi}$ is a provisional cutoff parameter 
(${\Lambda}_{\phi}^2 \gg m_{\phi}^2$),
which goes to the infinity in the limit of $m_{\phi}^2 \to 0$.
In case with ${\Lambda}_{\phi}^2 = (\Lambda^4/m_{\phi}^2)- m_{\phi}^2$,
we find that the second term on the right-hand side in (\ref{deltamphi-sep}) 
is invariant under the change $p^2 + m_{\phi}^2 \to \Lambda^4/(p^2 + m_{\phi}^2)$, 
but the first term is not.
Furthermore, in this case, the integration from $p^2=0$ to $p^2= {\Lambda}_{\phi}^2$ is divided into
that from $p^2=0$ to $p^2=\Lambda^2-m_{\phi}^2 (\ll {\Lambda}_{\phi}^2)$
and that from $p^2=\Lambda^2-m_{\phi}^2$ to $p^2= {\Lambda}_{\phi}^2$, 
and these integrals for the second term take a same value.
Note that the duality transformation reduces to that in the massless case,
in the limit of $m_{\phi}^2 \to 0$.

Here, we impose the duality relating $p^2 + m_{\phi}^2 \to \Lambda^4/(p^2 + m_{\phi}^2)$ 
on quantities relevant to $\Lambda$.
If the physics from $p^2=0$ to $p^2=\Lambda^2-m_{\phi}^2$ is 
same as that from $p^2=\Lambda^2-m_{\phi}^2$ to $p^2= {\Lambda}_{\phi}^2$
and the physics is only described by one of the two regions,
$\Lambda$ is naturally introduced and the desired expression is obtained as
\begin{eqnarray}
\delta m_{\phi}^2 = - \frac{\lambda_{\phi}m_{\phi}^2 }{32 \pi^2} 
\int_{0}^{\Lambda^2-m_{\phi}^2} \frac{dp^2}{p^2 + m_{\phi}^2}
= -\frac{\lambda_{\phi}}{32\pi^2} m_{\phi}^2 \ln \frac{\Lambda^2}{m_{\phi}^2}~.
\label{deltamphi-sep-Du}
\end{eqnarray}
Note that $\delta m_{\phi}^2$ vanishes in the limit of $m_{\phi}^2 \to 0$.

~~\\
(ii) The proper time method

Using the proper time method,
$\delta m_{\phi}^2$ is given as
\begin{eqnarray}
\delta m_{\phi}^2 
= \frac{\lambda_{\phi}}{2} \int_{-\infty}^{\infty} \frac{d^4p}{(2\pi)^4}
\int_0^{\infty} e^{- (p^2 + m_{\phi}^2) t} dt
= \frac{\lambda_{\phi}}{32 \pi^2} \int_0^{\infty} \frac{e^{- m_{\phi}^2 t}}{t^2} dt~,
\label{deltamphi-t}
\end{eqnarray} 
where $t$ is a parameter called a proper time.

First, we separate $\delta m_{\phi}^2$
into the quadratic and logarithmic divergent parts
by expanding the exponential factor such that
\begin{eqnarray}
\delta m_{\phi}^2 
= \frac{\lambda_{\phi}}{32 \pi^2} \int_{1/\tilde{\Lambda}_{\phi}^2}^{1/m_{\phi}^2} \frac{dt}{t^2}
- \frac{\lambda_{\phi}m_{\phi}^2 }{32 \pi^2} \int_{1/\tilde{\Lambda}_{\phi}^2}^{1/m_{\phi}^2} \frac{dt}{t}
+ \frac{\lambda_{\phi}m_{\phi}^4 }{64 \pi^2} \int_{1/\tilde{\Lambda}_{\phi}^2}^{1/m_{\phi}^2} dt
+ \cdots~,
\label{deltamphi-t-sep}
\end{eqnarray}
where $\tilde{\Lambda}_{\phi}$ is a provisional cutoff parameter
$(\tilde{\Lambda}_{\phi}^2 \gg m_{\phi}^2)$,
which goes to the infinity in the limit of $m_{\phi}^2 \to 0$.
In case with $\tilde{\Lambda}_{\phi}^2 = \Lambda^4/m_{\phi}^2$, 
we find that the second term on the right-hand side in (\ref{deltamphi-t-sep}) 
is invariant under the change $t \to 1/(\Lambda^4 t)$, but others are not.
Furthermore, in this case,
the integration from $t=1/\tilde{\Lambda}_{\phi}^2$ to $t= 1/m_{\phi}^2$ is divided into
that from $t=1/\tilde{\Lambda}_{\phi}^2$ to $t= 1/\Lambda^2$
and that from $t=1/\Lambda^2$ to $t= 1/m_{\phi}^2$, 
and these integrals for the second term take a same value.

Here, we impose the duality relating $t \to 1/(\Lambda^4 t)$ on quantities relevant to $\Lambda$.
If the physics from $t=1/\tilde{\Lambda}_{\phi}^2$ to $t= 1/\Lambda^2$ is 
same as that from $t=1/\Lambda^2$ to $t= 1/m_{\phi}^2$ 
and the physics is only described by one of the two regions,
$\Lambda$ is naturally introduced and the desired expression is obtained as
\begin{eqnarray}
\delta m_{\phi}^2 
= {\rm Du}\left[\frac{\lambda_{\phi}}{32 \pi^2} \int_0^{\infty} \frac{e^{- m_{\phi}^2 t}}{t^2} dt\right]
= - \frac{\lambda_{\phi}m_{\phi}^2 }{32 \pi^2} \int_{1/{\Lambda}^2}^{1/m_{\phi}^2} \frac{dt}{t}
= -\frac{\lambda_{\phi}}{32\pi^2} m_{\phi}^2 \ln \frac{\Lambda^2}{m_{\phi}^2}~.
\label{deltamphi-t-Du}
\end{eqnarray} 
In a similar way as the momentum cutoff method, $\delta m_{\phi}^2$ vanishes in the massless limit.\\

The region around $p^2={\Lambda}_{\phi}^2$ or $t=1/\tilde{\Lambda}_{\phi}^2$
corresponds to the ultra-violet (UV) one,
and that around $p^2=0$ or $t=1/m_{\phi}^2$ corresponds to the infra-red (IR) one.
Hence, the symmetry relating $p^2 + m_{\phi}^2 \to \Lambda^4/(p^2 + m_{\phi}^2)$ 
or $t \to 1/(\Lambda^4 t)$ might suggest 
that $\Lambda$ has a physical meaning as a fundamental scale and 
there is an equivalence between the physics at the UV region
and that at the IR one, in an ultimate theory.

\subsection{Different choice}
\label{Different choice}

It is important to examine the applicable scope of our scheme.
Here, we point out that the result depends on the choice of duality transformation,
by giving an example.

Based on the proper time method, 
$\delta m_{\phi}^2$ is rewritten as
\begin{eqnarray}
\delta m_{\phi}^2 = 
\frac{\lambda_{\phi}}{32 \pi^2} \int_0^{\infty} \frac{e^{- m_{\phi}^2 t}}{t^2} dt
= \frac{\lambda_{\phi}}{32 \pi^2} \int_0^{\infty} d \tau_2 \int_{-1/2}^{1/2} d \tau_1
\frac{\Lambda^2}{\tau_2^2}
e^{- \frac{m_{\phi}^2}{\Lambda^2} \tau_2}~,
\label{deltamphi-tau}
\end{eqnarray}
where $\tau_2 = \Lambda^2 t$.
Let us make the complex parameter
$\tau = \tau_1 + i \tau_2$ play the role of the modular parameter in string theory.
The world-sheet modular transformation is given by
\begin{eqnarray}
\tau \to \frac{a \tau + b}{c \tau + d}~,~~~ (ad-bc=1)
\label{modular-transf}
\end{eqnarray}
where $a$, $b$, $c$ and $d$ are integers,
and the transformation is generated by the compositions of two types of transformations
$\tau \to \tau + 1$ and $\tau \to -1/\tau$.
If we require the invariance under the transformation (\ref{modular-transf})
and assume that the physics is only described by an independent region, 
which is not connected with by the transformation,
the following expression is obtained,
\begin{eqnarray}
\delta m_{\phi}^2 = {\rm Du}\left[\frac{\lambda_{\phi}}{32 \pi^2} \int_0^{\infty} 
d \tau_2 \int_{-1/2}^{1/2} d \tau_1 \frac{\Lambda^2}{\tau_2^2}
e^{- \frac{m_{\phi}^2}{\Lambda^2} \tau_2}\right]
= \frac{\lambda_{\phi}}{32 \pi^2} \Lambda^2 \int_{\mathcal{F}} \frac{d^2\tau}{\tau_2^2}
= \frac{\lambda_{\phi}}{32 \pi^2} \frac{\pi}{2} \Lambda^2~,
\label{deltamphi-tau-Du}
\end{eqnarray}
where $\mathcal{F}$ stands for the fundamental region defined by
\begin{eqnarray}
\mathcal{F} = \{\tau: |\mbox{Re}\tau| \le {1}/{2}, 1 \le |\tau|\}~.
\label{F}
\end{eqnarray} 
The value of (\ref{deltamphi-tau-Du}) is different from that of (\ref{deltamphi-t-Du}).
The difference of values comes from that of the invariant measures, 
i.e., the invariant measure for $\tau \to -1/\tau$ is $d^2\tau/\tau_2^2$, but
that for $t \to 1/(\Lambda^4 t)$ is $dt/t$ up to a sign factor.
Note that both $\tau \to -1/\tau$ and $t \to 1/(\Lambda^4 t)$
connect the UV region to the IR one, and
$\tau \to -1/\tau$ reduces to $\tau_2 \to 1/\tau_2$, 
which corresponds to $t \to 1/(\Lambda^4 t)$, in case with $\tau_1 = 0$.

In this way, we find that the value of $\delta m_{\phi}^2$ depends on 
the form of duality transformation and need to specify it 
in order to obtain a physically meaningful value.
We expect that the form of duality transformation is determined
by matching the counterpart in the ultimate theory.

We add a comment on radiative corrections in string theory.
From the world-sheet modular invariance for the closed string,
$\delta m_{\phi}^2$ (radiative corrections of the scalar mass 
including contributions from innumerable string states)
should be given by
\begin{eqnarray}
\delta m_{\phi}^2 =
\int_{\mathcal{F}} \frac{d^2\tau}{\tau_2^2} G(\tau)~,
\label{deltamphi-st}
\end{eqnarray}
where $G(\tau)$ is a world-sheet modular invariant function, i.e., $G(\tau) = G(\tau+1)$
and $G(\tau) = G(-1/\tau)$.
In cases that SUSY holds exactly, $G(\tau)$ vanishes, and then $\delta m_{\phi}^2 = 0$.
Even if SUSY is broken down, there is a possibility that $G(\tau)$ vanishes in conspiracy with
infinite towers of massive particles,
as suggested in Ref.~\cite{Dienes}.

In string theory, the world-sheet modular invariance is deeply connected to the consistency
of theory, and radiative corrections should be given in the world-sheet modular invariant form
for the closed string.
On the other hand, in the effective field theory, a corresponding symmetry
stays in the background if it exists at all,
and the consistency of theory would not be necessarily threatened if it is overlooked.
Hence, radiative corrections in the effective theory
are not generally given in the duality invariant form, 
and the projection to pick out the invariant parts would be required.

Finally, we give a conjecture on a solution 
for the one side of the naturalness problem {\it whether the weak scale is stabilized 
against radiative corrections},
taking string theory as a candidate of the ultimate theory.
In string theory, the world-sheet conformal invariance induces the massless string states,
and the world-sheet modular invariance guarantees the finiteness of physical quantities.
We conjecture that, owing to some powerful symmetry (such as SUSY) 
in addition to the world-sheet modular invariance,
the masslessness of scalar particles would be protected against quantum corrections,
and the above-mentioned problem would not be caused.
This could be understood in the framework of low-energy effective field theory, as follows.
We assume that the theory is described by only massless string states,
effects of massive string states are introduced as non-renormalizable interactions
among massless particles, and they do not cause (the technical side of) the gauge hierarchy problem.
In the field theory limit, if $\tau$ reduces to $\tau_2$ with $\tau_1 =0$,
the duality transformation $\tau \to -1/\tau$ reduces to $\tau_2 \to 1/\tau_2$, 
which corresponds to $t \to 1/(\Lambda^4 t)$.
Then, massless scalar fields do not receive
any radiative corrections on their masses, as seen from (\ref{deltamphi-t-Du}). 
This matches the conjecture based on string theory.
For the case with massive light scalar fields, 
a more careful consideration is needed
and it is beyond the scope of this paper,
because it is deeply related to the other side of the naturalness problem
{\it what is the origin of the weak scale or the Higgs mass}.

\section{Conclusions}
\label{Conclusions}

We have reconsidered the naturalness and its relevant symmetries
from the viewpoint of effective field theories including the SM,
in the expectation that the unnaturalness for scalar masses
might be an artifact in the effective theory
and they could be improved if features of an ultimate theory are taken in
and ingredients of the effective theory are enriched.
We have given a suggestion for the subtraction of quadratic divergences,
based on the assumptions relating features of the ultimate theory.
The assumptions are summarized as follows.
Beyond the SM, there is an ultimate theory with a fundamental scale $\Lambda$
and a duality between the physics at the UV region beyond $\Lambda$ and that at the IR region,
and a remnant of the duality is hidden 
in the lower-energy theory.
We have shown that the logarithmic corrections can be picked out 
by specifying the duality transformation.
Because the logarithmic corrections are compatible with a specific duality, 
it is expected that the subtraction of quadratic divergences could be justified in the ultimate theory.

If the quadratic divergences of scalar fields are
artifacts of regularization procedure, 
the problem {\it whether the weak scale is stabilized 
against radiative corrections in the framework of SM}
can be solved by the subtraction of quadratic divergences.
Note that, even if the quadratic divergences are eliminated,
the physics beyond the SM can induce the gauge hierarchy problem, i.e.,
{\it the effective field theory becomes unnatural,
because a fine tuning is required to obtain the weak scale 
and/or to stabilize the weak scale, 
if there is a high-energy physics relevant to the SM.}
The sources of large radiative corrections, which can ruin the stability of the weak scale, 
are logarithmic divergences due to heavy particles.
There is a possibility that the SM (or the extension of the SM with new particles
around the terascale and without new concepts such as SUSY, compositeness
and extra dimensions) holds until $\Lambda$
and an ultimate theory protects masses of low-energy fields 
against large quantum corrections by some mechanism and/or symmetry.
This is a background of our previous work~\cite{K}.

Finally, we discuss the applicable scope of our method.
The issue is whether our calculation scheme is applicable to other systems
and ordinary results are obtained or not.
We anticipate that it is applicable 
to calculate logarithmic corrections including $\Lambda$ on quantities.

We have applied our method to the radiative corrections on vacuum energy density,
and shown that the logarithmic corrections can be picked out, in the appendix A.
Under the assumption that the QED holds until $\Lambda$ in the broken phase of electroweak symmetry,
we also have applied it to the self-energy of electron,
and obtained the well-known result, in the appendix B.

Because our procedure contains a provisional cutoff parameter depending on a mass,
it looks like an artifact or a temporary expedient to pick out specific corrections.
It is important to examine whether it is applicable to radiative corrections 
with higher loops by introducing several proper times 
and more complex models including several fields.

Even if our scheme has a limit of application or the hidden duality is a product of fantasy,
our expectation would survive that the calculation scheme can be selected by the physics,
and radiative corrections can be constrained
by a remnant of symmetries in an ultimate theory.

\section*{Acknowledgments}
This work was supported in part by scientific grants from the Ministry of Education, Culture,
Sports, Science and Technology under Grant Nos.~22540272 and 21244036.

\appendix
\section{Radiative corrections on vacuum energy density}

We apply our procedure to the radiative corrections on vacuum energy density $\delta \Lambda_{\rm V}$.
For the scalar field $\phi$, $\delta \Lambda_{\rm V}$ at the one-loop level is commonly written as
\begin{eqnarray}
&~& \delta \Lambda_{\rm V}
= - \frac{1}{2} \int_{-\infty}^{\infty} \frac{d^4p}{(2\pi)^4} \ln(p^2 + m_{\phi}^2)
= \frac{1}{2} \int_{-\infty}^{\infty} \frac{d^4p}{(2\pi)^4} \int_0^{\infty} \frac{e^{-(p^2 + m_{\phi}^2)t}}{t} dt
\nonumber \\
&~& ~~~~~~~~ 
= \frac{1}{32 \pi^2} \int_0^{\infty} \frac{e^{-m_{\phi}^2 t}}{t^3} dt~,
\label{deltaLambda-phi}
\end{eqnarray}
and it contains infinities.
We carry out the same procedure as that for scalar masses.

First, we separate $\delta \Lambda_{\rm V}$ 
into the quartic, quadratic and logarithmic divergent parts 
by expanding the exponential factor such that
\begin{eqnarray}
\hspace{-1.3cm}&~& \delta \Lambda_{\rm V}
= \frac{1}{32 \pi^2} \int_{1/\tilde{\Lambda}_{\phi}^2}^{1/m_{\phi}^2} \frac{dt}{t^3}
- \frac{1}{32 \pi^2} m_{\phi}^2 \int_{1/\tilde{\Lambda}_{\phi}^2}^{1/m_{\phi}^2} \frac{dt}{t^2}
\nonumber \\
\hspace{-1.3cm}&~& ~~~~~~~~~~
+ \frac{1}{64 \pi^2} m_{\phi}^4 \int_{1/\tilde{\Lambda}_{\phi}^2}^{1/m_{\phi}^2} \frac{dt}{t}
- \frac{1}{128 \pi^2} m_{\phi}^6 \int_{1/\tilde{\Lambda}_{\phi}^2}^{1/m_{\phi}^2} dt
+ \cdots~,
\label{deltaLambda-phi-t-sep}
\end{eqnarray}
where $\tilde{\Lambda}_{\phi}$ is a provisional cutoff parameter
$(\tilde{\Lambda}_{\phi}^2 \gg m_{\phi}^2)$,
which goes to the infinity in the limit of $m_{\phi}^2 \to 0$.
In case with $\tilde{\Lambda}_{\phi}^2 = \Lambda^4/m_{\phi}^2$, 
we find that the third term in the right-hand side of (\ref{deltaLambda-phi-t-sep}) 
is invariant under the change $t \to 1/(\Lambda^4 t)$, but others are not.

Here, we impose the duality relating $t \to 1/(\Lambda^4 t)$ on quantities relevant to $\Lambda$.
If the physics from $t=1/\tilde{\Lambda}_{\phi}^2$ to $t= 1/\Lambda^2$ is 
same as that from $t=1/\Lambda^2$ to $t= 1/m_{\phi}^2$
and the physics is only described by one of the two regions,
we obtain the relation
\begin{eqnarray}
\delta \Lambda_{\rm V}
= {\rm Du}\left[\frac{1}{32 \pi^2} \int_0^{\infty} \frac{e^{-m_{\phi}^2 t}}{t^3} dt\right]
= \frac{1}{64 \pi^2} m_{\phi}^4 \int_{1/{\Lambda}^2}^{1/m_{\phi}^2} \frac{dt}{t}
= \frac{1}{64 \pi^2} m_{\phi}^4 \ln \frac{\Lambda^2}{m_{\phi}^2}~.
\label{deltaLambda-phi-t-reg-st}
\end{eqnarray}

In this way, the quartic and quadratic divergences in $\delta \Lambda_{\rm V}$ are 
eliminated by requiring that the effective theory should have a hidden symmetry on the proper time.
Note that $\delta \Lambda_{\rm V}$ also vanishes,
in the massless limit $m_{\phi} =0$.
Because the subtraction of the quartic and quadratic divergences in $\delta \Lambda_{\rm V}$ 
induces the effect that the cosmological constant shifts,
a more careful consideration is required to justify our procedure.\footnote{
As another work to show the importance of the trans-Planckian physics,
Volovik gave the observation that 
the sub-Planckian and trans-Planckian contributions to 
the vacuum energy are canceled by the thermodynamical argument~\cite{Volovik}.
}

\section{Self-energy of electron}

The self-energy of electron with the momentum $p$ at the one-loop level is given by~\cite{B&D}
\begin{eqnarray}
\varSigma(p) 
= - e^2 \int_{-\infty}^{\infty} \frac{d^4q}{(2\pi)^4} \left(\frac{i}{q^2 + \mu_{\gamma}^2}  \gamma_{\nu}
\frac{i}{\Slashp - \Slashq - m_{\rm e}} \gamma^{\nu}\right)~,
\label{Sigma}
\end{eqnarray}
where we rotate to the Euclidean space,
$q$ is a momentum of virtual photon, $p-q$ is a momentum of virtual electron,
and $\mu_{\gamma}$ is a fictitious photon mass
for a regularization of IR divergences occurring $q^2=0$.
Using the proper time method, $\varSigma(p)$ is written by
\begin{eqnarray}
&~& \varSigma(p) 
= e^2 \int_0^{\infty} dz_1 \int_0^{\infty} dz_2 \int_{-\infty}^{\infty} \frac{d^4q}{(2\pi)^4} 
\left[\gamma_{\nu} \left(\Slashp - \Slashq + m_{\rm e}\right) \gamma^{\nu} \right]
\nonumber \\
&~& ~~~~~~~~~~ \times \exp\left[-z_1\left(q^2 + \mu_{\gamma}^2\right)
- z_2\left((p-q)^2 + m_{\rm e}^2\right)\right]~.
\label{Sigma2}
\end{eqnarray}
By changing the integration variable $q$ into the following one
\begin{eqnarray}
\tilde{q} \equiv q -\frac{z_2}{z_1 + z_2} p = q - p + \frac{z_1}{z_1 + z_2} p
\label{tildeq}
\end{eqnarray}
and integrating out $\tilde{q}$,
we obtain the expression
\begin{eqnarray}
&~& \varSigma(p) 
= \frac{e^2}{16\pi^2} 
\int_0^{\infty} \int_0^{\infty} \frac{dz_1 dz_2}{(z_1+z_2)^2}
\left(- 2\frac{z_1}{z_1+z_2}\Slashp + 4m_{\rm e} \right)
\nonumber \\
&~& ~~~~~~~~~~ \times \exp\left[-\left(\frac{z_1 z_2}{z_1+z_2}p^2 + z_1 \mu_{\gamma}^2
+ z_2 m_{\rm e}^2\right) \right]~.
\label{Sigma4}
\end{eqnarray}
Furthermore, we insert the following identity relating the delta function into the integrand,
\begin{eqnarray}
\int_0^{\infty}  d\xi~\delta(\xi - z_1 - z_2)=
\int_0^{\infty} \frac{d\xi}{\xi} \delta\left(1-\frac{z_1+z_2}{\xi}\right) = 1~,
\label{delta0}
\end{eqnarray}
and integrate out $z_2$ after changing the scale of the proper time parameters $z_i$ as $\xi z_i$,
we obtain the expression
\begin{eqnarray}
\varSigma(p)
= \frac{e^2}{8\pi^2} \int_0^{1} dz_1
\left(- z_1 \Slashp + 2m_{\rm e} \right)
\int_0^{\infty} \frac{d\xi}{\xi} e^{-\xi \tilde{m}^2}~,
\label{Sigma5}
\end{eqnarray}
where $\tilde{m}^2$ is a function of $p^2$ and $z_1$, defined by
\begin{eqnarray}
\tilde{m}^2 \equiv z_1(1-z_1)p^2 + z_1 \mu_{\gamma}^2 + (1-z_1) m_{\rm e}^2~.
\label{tildem}
\end{eqnarray}

We expand the exponential factor such that
\begin{eqnarray}
&~& \varSigma(p)
= \frac{e^2}{8\pi^2} \int_0^{1} dz_1
\left(- z_1 \Slashp + 2m_{\rm e} \right)
\int_{1/\tilde{\Lambda}_{p}^2} ^{1/\tilde{m}^2} \frac{d\xi}{\xi}
\nonumber \\
&~& ~~~~~~~~~ 
- \frac{e^2}{8\pi^2} \int_0^{1} dz_1
\left(- z_1 \Slashp + 2m_{\rm e} \right) \tilde{m}^2
\int_{1/\tilde{\Lambda}_{p}^2} ^{1/\tilde{m}^2} d\xi
+ \cdots~,
\label{Sigma-reg}
\end{eqnarray}
where $\tilde{\Lambda}_{p}$ is a provisional cutoff parameter
$(\tilde{\Lambda}_{p}^2 \gg \tilde{m}^2)$.
In case with $\tilde{\Lambda}_{p}^2 = z_1^2 \Lambda^4/\tilde{m}^2$,
we find that the first term in the right-hand side of (\ref{Sigma-reg}) 
is invariant under the change $\xi \to 1/(z_1^2 \Lambda^4 \xi)$, but others are not.
Furthermore, in this case,
the integration from $\xi=\tilde{m}^2/z_1^2 \Lambda^4$ to $\xi= 1/\tilde{m}^2$ is divided into
that from $\xi=\tilde{m}^2/z_1^2 \Lambda^4$ to $\xi = 1/(z_1 \Lambda^2)$
and that from $\xi =1/(z_1 \Lambda^2)$ to $\xi= 1/\tilde{m}^2$, 
and these integrals for the first term take a same value.
If we identify $z_1 \xi$ as a proper time $t$,
the transformation is the same form $t \to 1/(\Lambda^4 t)$ 
as that in case of the scalar mass $m_{\phi}$.

Here, we impose the duality relating $\xi \to 1/(z_1^2 \Lambda^4 \xi)$ on quantities relevant to $\Lambda$.
If the physics from $\xi=\tilde{m}^2/z_1^2 \Lambda^4$ to $\xi = 1/(z_1 \Lambda^2)$
is same as that from $\xi =1/(z_1 \Lambda^2)$ to $\xi= 1/\tilde{m}^2$
and the physics is only described by one of the two regions,
the desired expression is obtained as
\begin{eqnarray}
&~& \varSigma(p)
= \frac{e^2}{8\pi^2} \int_0^{1} dz_1
\left(- z_1 \Slashp + 2m_{\rm e} \right)
\int_{1/(z_1 {\Lambda}^2)} ^{1/\tilde{m}^2} \frac{d\xi}{\xi}
\nonumber \\
&~& ~~~~~~~~~
= \frac{e^2}{8\pi^2}\int_0^1 \left(- z_1 \Slashp + 2m_{\rm e}\right)
\ln\frac{z_1 \Lambda^2}{(1-z_1)m_{\rm e}^2 + z_1 \mu_{\gamma}^2 + (1-z_1)z_1 p^2} dz_1~.
\label{Sigma-reg2}
\end{eqnarray}
Using (\ref{Sigma-reg2}), we obtain the ordinary expression for radiative corrections 
on the electron mass such that
\begin{eqnarray}
\delta m_{\rm e} 
= \left. \varSigma(p)\right|_{\Slashp =m_{\rm e}}
= \frac{e^2 m_{\rm e}}{8\pi^2}\int_0^1 (2 - z_1) \ln\frac{z_1 \Lambda^2}{(1-z_1)^2m_{\rm e}^2} dz_1
= \frac{3\alpha}{4\pi} m_{\rm e} \left(\ln \frac{\Lambda^2}{m_{\rm e}^2}+ \frac{1}{2}\right)~,
\label{deltame-reg}
\end{eqnarray}
where we take the limit of $\mu_{\gamma}^2 \to 0$.

\end{document}